\documentclass[12pt]{article}
\usepackage{theorem,amssymb,amsbsy,latexsym,amsmath,bbm,epsfig,psfrag}

\theoremstyle{plain}

\newcommand{\Mpl}{M_{Pl}}
\newcommand{\define}{\equiv}

\newcommand{\Ref}[1]{(\ref{#1})}
\newcommand{\R}{{\mathbb R}}

\begin{document}
\begin{flushright}
March 25, 2011

\end{flushright}
\vspace{.4cm}

\begin{center}
  
\vspace{1 cm}
{\Large \bf Extrinsic curvature effects in brane-world scenarios}\\
[2mm] 
{\bf Edwin Langmann and Martin Sundin} \\
[2mm] 
{\it Theoretical Physics, KTH\\
SE-10691 Stockholm, Sweden}\\ 
{\tt langmann@kth.se, masundi@kth.se} \\
[5mm]
\end{center}

\begin{abstract}
We consider models of bosons on curved 3+1 dimensional space-time  embedded in a higher dimensional flat ambient space. We propose to derive (rather than postulate) equations of motions by assuming that a standard Klein-Gordon field on ambient space is restricted to space-time by a strong confining potential.  This leads to a modified Klein-Gordon equation on space-time which includes, in addition to the standard terms, a term with a so-called {\em induced potential} which depends on intrinsic- and extrinsic curvature  of the embedded space-time but not on the details of the confining potential. We compute this induced potential for natural, simple embeddings of Schwarzschild- and Robertson-Walker space-times. We also discuss possible observable implications of our results and, in particular, propose and study an extension of a standard model of cosmological inflation taking into account extrinsic curvature effects. We show that  the modified model allows for a solution where the scaling function vanishes like a power law with exponent $(\sqrt{3}+1)/4\approx 0.683$ at some initial time.
\end{abstract}

\section{Introduction}
\label{sec1}
A fundamental postulate in Einstein's theory of gravitation is that physics only depends  on intrinsic geometric properties of  curved space-time.   Many approaches to gravity and cosmology assume that space-time is a brane, i.e., a submanifold embedded in a higher dimensional ambient space \cite{Maartens} (see \cite{Branes} for an extensive list of commented references).  In such a brane world scenario, Einstein's postulate cannot be assumed  but rather is a property that should be  derived from an underlying theory.  One intriguing possibility is that this property holds true only in some approximation.  In this case one might be able to predict physical effects that depend on extrinsic curvature. This would be   interesting as a  means to falsify theoretical proposals. It also might suggest experiments that give direct evidence for (or against) the existence of extra dimensions.  In this paper we propose a possible  cause for extrinsic curvature effects that, as we believe, is potentially relevant in brane world scenarios. We also discuss possible physical implications. Our proposal is motivated by well-established results on constrained quantum mechanics, as discussed in the next paragraph. 

As proposed already by Schr\"odinger in 1926, it is natural to postulate that the  quantum mechanical Hamiltonian for a free particle  on a submanifold of three dimensional Euclidean space is proportional to the Laplace-Beltrami operator on this submanifold, and it thus only depends on intrinsic geometry;  see Equation~(31) in \cite{Schrodinger}. A more physical method is to derive  this Hamiltonian as follows: use the well-established Hamiltonian in three dimensional space and restrict the particle to the manifold  by a strong confining potential. As discovered by Jensen and Koppe \cite{JensenKoppe} in this context,\footnote{A similar result was found earlier by Marcus \cite{Marcus} in an effective quantum model for chemical reactions.} the effective Hamiltonian thus obtained contains, in addition to the expected Laplace-Beltrami operator term, an induced potential that depends on intrinsic- and extrinsic curvature; see \cite{C1,C2,MD,Mitchell,FH,SJ,Gol,WT} for generalizations and alternative derivations of this result. For example, the induced potential for a curve is attractive and proportional to the curvature. This explains why, as shown by Exner and Seba \cite{ES}, an electron on a wire has bound states when the wire has non-zero curvature;\footnote{Note that the intrinsic geometry of a curve is trivial, and thus Schr\"odinger's postulated effective Hamiltonian cannot explain such bound states.}  see also \cite{GJ} for similar results in the context of electromagnetic waveguides.  The existence of the induced potential was confirmed experimentally; see e.g.\ \cite{CLMTY} and references therein. 

Consider a real-valued Klein-Gordon field $\phi$ on curved space-time embedded in higher dimensional Minkowski space. We propose to {\em derive}  an effective equation of motion for $\phi$ from a Klein-Gordon field on ambient space restricted to the space-time submanifold by a strong confining potential. As we show, this yields an effective description by the following modified Klein-Gordon equation on space-time\footnote{Our notation is explained in the next section. }
\begin{equation}
\label{KG1} 
\Bigl( \hbar^2\Bigl[ |g|^{-1/2} \partial_\mu|g|^{1/2}g^{\mu\nu}\partial_\nu  + V_{ind}(x)\Bigr]+ (m_0 c)^2 \Bigr)\phi(x) =0
\end{equation} 
with the induced potential $V_{ind}$ in Equation~\Ref{Vind} below (we use Planck's constant $\hbar$ and the vacuum velocity of light $c$ only here, to emphasize that the induced potential is a correction to the Laplace-Beltrami operator;  in the rest of the paper we set $\hbar=c=1$). In fact, our result is more general: we use the Lagrangian formalism, and \Ref{KG1} corresponds to the Euler-Lagrange equations of the non-interacting part of the action only, but our result applies also to the case with interactions. It is worth noting that  $V_{ind}$  is universal in the sense that it is independent of the details of the confining potential. We also investigate if and when   $V_{ind}$ could lead to measurable effects. For that we compute  $V_{ind}$ for various special cases.

We emphasize that, for given space-time, the embedding in a higher dimensional Minkowski space is not unique, and the induced potential depends on the embedding. Our proposal can therefore only lead to a definite prediction in combination with a theory that fixes the embedding. We do not assume such a theory but instead use simple embeddings in flat ambient spaces with the lowest possible dimension. Fortunately, in many examples of physical interest (including the ones we discuss), there exists one such embedding which is preferred by its naturalness and simplicity; see \cite{Rosen} for a list of known examples. We thus believe that our approach is justified by "Ockham's razor". The investigation of more complicated embeddings (motivated by string theory etc.) is left to future work. 

While the formula for the induced potential we obtain is identical with the obvious generalization of the well-known one in constrained quantum mechanics \cite{JensenKoppe,C2,MD} from Riemannian- to pseudo-Riemannian spaces, the physics is very different, and the derivations of the induced potential in the literature do not apply to our case. We therefore present a different derivation for pseudo-Riemannian manifolds $\mathcal{M}^{q,n}$ of arbitrary signatures $(q,n)$ embedded in flat ambient spaces $\mathbb{R}^{q,n+p}$ (i.e., the number $q\geq 0$ and $n\geq 1$ of time- and space-like dimensions in $\mathcal{M}^{q,n}$ is arbitrary, and so is the number $p\geq 1$ of the extra dimensions). To not further burden our notation, we restrict our derivation to the case $q=1$ of main interest to us, but the generalization to arbitrary $q$ is obvious. To be more specific: Derivations of the induced potential in constrained quantum mechanics are usually based on equations of motion; see \cite{WT}, Section~I.A for a more detailed discussion of the history and the different levels of mathematical rigor of these derivations. A key point  in such quantum mechanical derivations is the probability interpretation of the quantum mechanical wave function. This provides an argument to use a scaling factor for the wave function in ambient space, and this scaling factor leads to the induced potential; see e.g.\ \cite{SJ}, Equation~(12) {\em ff} for a lucid discussion of this point. However, Klein-Gordon fields do not have such a probability interpretation, and this argument therefore cannot be used. Instead, we start with the standard action for a Klein-Gordon field on flat ambient space and with a suitable strong confining potential; see Equation~\Ref{S0} below. By expanding this field in suitable modes and straightforward computations, we find that this action describes coupled Klein-Gordon fields on space-time; see Equation~\Ref{action2}. Using a standard physics argument, we finally reduce the latter to an effective action for a single Klein-Gordon fields; see Equation~\Ref{Seff}. The above-mentioned scaling factor arises in this derivation for purely mathematical reasons. Our derivation suggests that such an effective action cannot be obtained in cases where the number of time-like directions in physical space-time and ambient space are different.
 
In the rest of this paper we first introduce  our notation and present our derivation of the induced potential for confined Klein-Gordon fields (Section~\ref{sec2}). Our examples are in Sections~\ref{sec3}--\ref{sec6}: We first give the induced potential for the Schwarzschild metric and discuss its possible physical implications for black holes (Section~\ref{sec3}).  We then consider the induced potential for the closed Robertson-Walker metric (Section~\ref{sec5}), and we use this to construct and study a brane-world model of cosmological inflation taking into account extrinsic curvature effects (Section~\ref{sec6}). We end with conclusions in Section~\ref{sec7}. Some details of our computations are given in two appendices. 

We will mention basic results on general relativity and cosmology that are discussed in several textbook (including \cite{Borner,LythLiddle09}) without further reference. 

\section{General result}
\label{sec2}
We assume that space-time $\mathcal{M}\equiv \mathcal{M}^{1,n}$ is a $(n+1)$-dimensional Lorentzian manifold embedded in $(n+p+1)$-dimensional Minkowski space $\R^{1,n+p}$. 

\noindent {\bf Notation:}  We use capital latin letters $M,N,\dots$ for indices running over $0,1,2,\dots,n+p$, and denote as $\mathbf{Z}\equiv(Z^0,Z^1,Z^2,\dots,Z^{n+p})$ inertial coordinates in $\R^{1,n+p}$, i.e., the line element in these coordinates is (the following defines our sign conventions) 
\begin{equation}
\label{ds2} 
ds^2 = \eta_{MN}dZ^MdZ^N \equiv (dZ^0)^2 - (dZ^1)^2 - \dots - (dZ^{n+p})^2. 
\end{equation} 
Greek letters $\mu,\nu,\lambda,\sigma,\ldots$ and lower case latin letters $i,j,k\dots$ are used for indices running over $0,1,2,\dots,n$ and $n+1,n+2,\dots,n+p$, respectively. We write bold face letters for vectors in $\R^{1,n+p}$ and a dot for the scalar product of such vectors, i.e., $\mathbf{v}\cdot\mathbf{w}\equiv\eta_{MN} v^M w^N$. We assume that space-time $\mathcal{M}$  can be (locally) parametrized by a function $\mathbf{Z} = \mathbf{f}(x)$ with $x\equiv(x^0,\ldots,x^n)$ in an open subset of $\mathbb{R}^{1,n}$. The metric tensor of $\mathcal{M}$ is then $g_{\mu\nu}=\mathbf{t}_\mu \cdot \mathbf{t}_\nu$, with tangent vectors $\mathbf{t}_\mu(x) \equiv \partial_\mu\mathbf{f}(x)$.\footnote{We write $\partial_M$, $\partial_\mu$ and $\partial_i$ short for $\frac{\partial}{\partial Z^M}$, $\frac{\partial}{\partial x^\mu}$ and $\frac{\partial}{\partial y^i}$, respectively. We always assume implicitly that functions we introduce are differentiable up to the degrees needed.} Using $y\equiv (y^{n+1},\dots,y^{n+p}) \in \mathbb{R}^{p}$ we construct a local coordinate system of an open neighborhood of $\mathcal{M} \subset \mathbb{R}^{1,n+p}$ by setting
\begin{equation}
\label{tildeZ}
\mathbf{\tilde{f}}(x,y) = \mathbf{f}(x) + y^i \mathbf{n}_i(x) 
\end{equation} 
with $p$ linearly independent space-like vectors $\mathbf{n}_i$ orthogonal to all tangent vectors 
and with constant scalar products, i.e., 
\begin{equation}
\label{t and n}
\mathbf{t}_\mu (x) \cdot \mathbf{n}_i (x) =0,\quad \mathbf{n}_i (x) \cdot \mathbf{n}_j (x) = -h_{ij}
\end{equation} 
for all $x$, $\mu$ and $i$, with constant $h_{ij}$ defined by this equation. We can further restrict the vectors $\mathbf{n}_i$ by the additional condition $h_{ij}=\delta_{ij}$, without loss of generality. However, we sometimes find it convenient to leave $h_{ij}$ general in our equations. We assume that there exists $\epsilon>0$ such that the coordinate system $(x,y)$ is one-to-one for all $y$ such that $|y|\equiv \sqrt{h_{ij}y^iy^j}<\epsilon$ (as discussed in the beginning of Appendix~\ref{appA1}, we believe that this is a minor restriction). We denote the metric tensor on $\R^{1,n+p}$ in the coordinates $(x,y)$ by $G_{MN}(x,y)$. Here and in the following we write $|h|$, $|g|$ and $|G|$ for the absolute values of the determinants of $(h_{ij})$, $(g_{\mu\nu})$ and $(G_{MN})$, respectively. We define $G^{MN}$ as usual: $G^{ML}G_{LN}=\delta^M_{\phantom M N}$, and similarly for $g^{\mu\nu}$ and $h^{ij}$. 

We consider a Klein-Gordon field $\Phi\equiv \Phi(\mathbf{Z})$ on ambient space $\R^{1,n+p}$ with the usual dynamics but confined to the region close to the submanifold $\mathcal{M}$ by a potential $V_{conf}(\mathbf{Z})$, i.e., the (free part of)  the action is
\begin{equation}
\label{S0}
S_0=\frac12\int_{\R^{1,n+p}} d^{n+p+1}Z\Bigl(\eta^{MN}\bigl(\partial_M \Phi\bigr)\bigl(\partial_N\Phi\bigr) -(m_{bare}^2+V_{conf})\Phi^2  \Bigr)
\end{equation}
with the ``bare mass parameter'' $m_{bare}^2\in\R$. We assume that the potential $V_{conf}$ is strongly confining to $\mathcal{M}$, i.e., in the coordinates $(x,y)$ it has the form 
\begin{equation}
\label{Vconf} 
\tilde V_{conf}(x,y)\define
V_{conf}(\tilde{\mathbf{f}}(x,y))=\frac1{\epsilon^2} V(y/\epsilon)+ O(|y|)   
\end{equation}
with $\epsilon>0$ a convenient scaling parameter assumed to be small;\footnote{Introducing this scaling parameter is a useful mathematical  trick allowing to cleanly separate different energy scales; see e.g.\ \cite{WT} for a lucid discussion of this point.} we allow for a possible correction term $O(|y|)$ depending on $x$ but vanishing at least linearly with $|y|$ as $|y|\to 0$. We assume that $V(y)$ is such that the eigenvalue equation\footnote{To be more precise, we assume that $V(y)$ is such that the following is an eigenvalue equation of a self-adjoint Schr\"odinger operator on the Hilbert space of square integrable function on $\R^p$.}
\begin{equation}
\label{def chi}
-h^{ij}\partial_i\partial_j\chi_\alpha(y) + V(y)\chi_\alpha(y)=\mu_\alpha \chi_\alpha(y) 
\end{equation} 
has a unique solution $\chi_0$ corresponding to the smallest possible eigenvalue $\mu_0$ and such that $\int d^py\, |h|^{1/2}|y|^n|\chi_0(y)|^2$ is finite for all $n=0,1,2,\ldots$, and $\mu_\alpha-\mu_0>0$ for all $\alpha\neq 0$;  here and in the following, indices $\alpha,\beta\ldots$  denote quantum numbers labeling the solutions of the eigenvalue equation in \Ref{def chi}. Below we refer to such a potential $V(y)$ as {\em suitable}.  We choose the eigenfunctions $\chi_\alpha$ to be real-valued and normalized such that
\begin{equation}
\label{def chi 1} 
\int d^p y\, |h|^{1/2}\chi_\alpha(y)\chi_\beta(y)=\delta_{\alpha\beta}.  
\end{equation} 
To be specific we mention one example for a suitable potential allowing for simple computations of all solutions of \Ref{def chi}: $h_{ij}=\delta_{ij}$ and 
\begin{equation}
\label{box} 
V(y) = \lim_{v_0\to+\infty}v_0\Bigl(1 - \prod_{j=n+1}^{n+p}\theta(\ell-|y^j|) \Bigr) 
\end{equation} 
with $\ell>0$ and the Heaviside function $\theta$ (``infinite box potential'') but, as discussed in Appendix~\ref{appA1}, the same result is obtained for a large class of potentials. In particular, $V$ can be bounded and such that the eigenvalue equation in \Ref{def chi} also has scattering solutions (in this case the symbols $\sum_\alpha$ and $\delta_{\alpha\beta}$ have to be partly interpreted as integral and Dirac delta). 

One can expand $\tilde\Phi(x,y)\define\Phi(\mathbf{\tilde f}(x,y))$ in the eigenfunctions $\epsilon^{-p/2}\chi_\alpha(y/\epsilon)$ and thus rewrite the action in \Ref{S0} as an action of an infinite number of fields $\phi_\alpha(x)$ on space-time $\mathcal{M}$. As explained in Appendix~\ref{appA2}, a key point in this computation is that, to correct a mismatch of Jacobian determinants, one has to include a scaling factor $(|h||g|/|G|)^{1/4}$ in this expansion as follows, 
\begin{equation}
\label{expansion} 
\tilde\Phi(x,y)=\left(\frac{|h||g(x)|}{|G(x,y)|}\right)^{1/4}\sum_\alpha\phi_\alpha(x) \epsilon^{-p/2}\chi_\alpha(y/\epsilon). 
\end{equation} 
By a straightforward computation we obtain the following.

\noindent \textbf{Result:} {\em For suitable confining potentials $V$ and sufficiently small $\epsilon>0$, the action in \Ref{S0} equals
\begin{equation}
\label{action2}
S_0 =  \frac12\int_{\mathcal{M}} d^{n+1}x\, |g|^{1/2}\sum_\alpha \Biggl(g^{\mu\nu} \bigl(D_\mu\phi\bigr)_\alpha\bigl(D_\nu \phi\bigr)_\alpha - \bigl(m_{bare}^2 + \mu_\alpha/\epsilon^2+ V_{ind}\bigr)\phi_\alpha^2 \Biggr) + O(\epsilon)
\end{equation}
\begin{equation}
\label{Ddef}
\bigl(D_\mu\phi\bigr)_\alpha\equiv \partial_\mu\phi_\alpha + \sum_{\beta\neq \alpha }C_{\mu\alpha\beta}\phi_\beta 
\end{equation} 
with
\begin{equation}
\label{Cdef} 
C_{\mu\alpha\beta} = A_{j\mu}^{\phantom{j\nu}i} \int_{\R^p} d^py\, |h|^{1/2}  y^j \chi_\beta (y) \partial_i \chi_\alpha (y)
\end{equation}
and 
\begin{equation}
\begin{split}
\label{Vind}
V_{ind} =  \frac{1}{4} h^{ij}\left(\alpha_{i\lambda}^{\phantom{i\lambda}\lambda}\alpha_{j\sigma}^{\phantom{j\sigma}\sigma}- 2\alpha_{i\lambda}^{\phantom{i\lambda}\sigma}\alpha_{j\sigma}^{\phantom{j\sigma}\lambda}\right)
\end{split} 
\end{equation}
where $\alpha_{i\mu}^{\phantom{i\mu}\lambda}(x)$ and $A_{i\mu}^{\phantom{i\mu}k}(x)$ are defined by the following equation, 
\begin{equation}
\label{alpha and A}
\partial_\mu \mathbf{n}_i = - \alpha_{i \mu}^{\phantom{i\mu}\lambda} \mathbf{t}_\lambda - A_{i\mu}^{\phantom{i\mu}k} \mathbf{n}_k.
\end{equation}
} 

\noindent (See Appendix~\ref{appA} for a derivation of this result.) 

We note in passing that the action in \Ref{action2}--\Ref{Vind} has a gauge theory structure discussed in \cite{MD,SJ}, for example, but this is not used in the present paper. 

Thus, up to terms $O(\epsilon)$ vanishing in the limit $\epsilon\to 0^+$, the confined Klein-Gordon action on ambient space in \Ref{S0} is equivalent to an action of Klein-Gordon fields $\phi_\alpha$ on $\mathcal{M}$ with effective masses
\begin{equation}
m_\alpha^2 = m_{bare}^2 + \mu_\alpha/\epsilon^2 
\end{equation}
and additional potential- and derivative terms. It is natural to fix the renormalized mass $m_0^2$ and choose $m_{bare}^2=m_0^2-\mu_0/\epsilon^2$. The other mass parameters $m^2_{\alpha\neq 0}=m_0^2+(\mu_\alpha-\mu_0)/\epsilon^2$ are then much larger than $m_0^2$. Standard physics arguments suggest that, for sufficiently small values of $\epsilon$, only the Klein-Gordon field $\phi_0$ with the ``small'' mass parameter $m_0^2$ is relevant for the low energy physics properties of the model. It therefore is a good approximation to simplify the model by replacing the action in \Ref{action2} with 
\begin{equation}
\label{Seff} 
S^{(0)}_{eff} = \frac12\int_{\mathcal{M}} d^{n+1}x\, |g|^{1/2} \left(g^{\mu\nu} \bigl( \partial_\mu \phi\bigr)\bigl(\partial_\nu \phi\bigr) - (m^2_0 + V_{ind}) \phi^2 \right)
\end{equation} 
and $\phi\equiv \phi_0$. This is our result described in the introduction.  It is important that it is independent of the details of the confining mechanism: changing $V$ can be compensated by a change of the bare mass parameter $m_{bare}^2$ and thus is irrelevant. 

As already mentioned, the generalization of this result to interacting Klein-Gordon fields is straightforward. In particular, changing the action in \Ref{S0} by a $\Phi^4$-interaction term,  
\begin{equation}
S = S_0 - \frac{\lambda_{bare}}4\int_{\R^{r+p+1}}d^{n+p+1}Z\, \Phi^4, 
\end{equation}
leads to the following change of the effective action in \Ref{Seff}, 
 \begin{equation}
\label{Seff1} 
S_{eff} = S^{(0)}_{eff} - \frac{\lambda_0}4 \int_{\mathcal{M}} d^{n+1}x\, |g|^{1/2}\phi^4
\end{equation} 
with the renormalized interaction strength 
\begin{equation}
\lambda_0 = \lambda_{bare}\epsilon^{-p} \int_{\R^p}  d^p y\, |h|^{1/2}\chi_0(y)^4. 
\end{equation} 
Again it is natural to fix $\lambda_0>0$ and adapt $\lambda_{bare}$ accordingly, i.e., the effective action is independent of the details of the confining mechanism also in the presence of interactions. 

\section{The Schwarzschild black hole}
\label{sec3} 
The Schwarzschild metric is a spherically symmetric solution of Einstein's equations describing a black hole with mass $M$ in an otherwise empty universe. It is given by the line element
\begin{equation}
\label{ds2_Schwarzschild} 
ds^2 = \left( 1 - \frac{r_s}{r}\right) dt^2 - \left( 1 - \frac{r_s}{r}\right)^{-1} dr^2 - r^2  (d\theta^2 + \sin^2(\theta)d\varphi^2)
\end{equation}
where $r_s = 2GM$ is the Schwarzschild radius and $x=(t,r,\theta,\varphi)$ the Schwarzschild coordinates, as usual. 

As proved by Kasner \cite{K1}, the Minkowski space-time of lowest dimension allowing an embedding of Schwarzschild space-time is $\mathbb{R}^{1,5}$. One such embedding found by Fronsdal \cite{Fronsdal} is given by
\begin{equation} 
\begin{split}
\label{embed_Schwarzschild} 
  &Z^0 = \begin{cases} 2r_s(1-r_s/r)^{1/2} \sinh(t/[2r_s])& (r>r_s)\\ 2r_s(r_s/r-1)^{1/2} \cosh(t/[2r_s])& (0<r<r_s) \end{cases}\\   
&Z^1 = \begin{cases} 2r_s(1-r_s/r)^{1/2} \cosh(t/[2r_s])& (r>r_s)\\ 2r_s(r_s/r-1)^{1/2} \sinh(t/[2r_s])& (0<r<r_s) \end{cases}\\   
&Z^2 = g(r)\\
&Z^3 = r\sin(\theta)\cos(\varphi)\\
&Z^4 = r\sin(\theta)\sin(\varphi)\\
&Z^5 = r\cos(\theta)
\end{split}
\end{equation}
where 
\begin{equation}
\label{gprime}  
g'(r)\equiv \frac{dg(r)}{dr} = \sqrt{\frac{r_s(r^2+rr_s+r_s^2)}{r^3}}.
\end{equation} 
This embedding is natural in that no other complete embedding of the Schwarzschild solution in the six-dimensional Minkowski space-time exists \cite{Sassi}. 

By straightforward computations we find the induced potential (see Appendix~\ref{appB1} for details)
\begin{equation}
\label{Schwarzschild_result1}
V_{ind}(r) = -\frac{(\hat r^3+ \hat r^2+ \hat r+9)(\hat r^2+1)(\hat r+1)}{16r_s^2 \hat r^4(\hat r^2+\hat r+1)},\quad \hat r\equiv r/r_s
\end{equation}
which is non-singular in the whole region $0<r<\infty$. Considering that the details of the embedding are different for $r>r_s$ and $r<r_s$, it is remarkable that the induced potential is continuous at $r=r_s$. Note that $V_{ind}(r)$ is strictly monotone and has the following asymptotic behavior, 
\begin{equation}
\label{Schwarzschild_result2}
V_{ind}(r) =\begin{cases} -9/(16 r_s^2 \hat r^4)\Bigl( 1 + \hat r/9 +O(\hat r^2)\Bigr) & (r\to 0^+)\\
-1/r_s^2\Bigl(1-3(\hat r-1)+O\bigl((\hat r-1)^2\bigr)\Bigr)& (r\to r_s) \\
-1/(4r_s)^2\Bigl( 1 + 1/\hat r + O(1/\hat r^{2})  \Bigr) & (r\to \infty) \end{cases}. 
\end{equation}
Moreover, it approaches its limiting value $-1/(4r_s)^2$ rather rapidly and, for  distances $r$ larger than $100r_s$, $V_{ind}(r)$ differs from its limiting value by less than $1\%$. Note that the magnitude of this limiting value corresponds to a boson mass, in SI units, 
\begin{equation}
\label{hbar}
\frac{\hbar}{4 c r_s}  \approx  \frac{3.31\times 10^{10}\, \mathrm{kg}}{M}\frac{\mathrm{GeV}}{c^2}.
\end{equation} 

We conclude this section with a short discussion of the possible physical relevance of our result.  It is worth noting that the boson-mass equivalent of the magnitude of the induced potential in \Ref{hbar}  is proportional to the {\em inverse} of the black hole mass $M$. Moreover, it seems that only very light black holes, with masses $10^{10}\,$kg or smaller, could give rise to induced potentials that have measurable effects, and the spatial variation of this potential is significant only very close to such a black hole. Thus the natural candidates to search for such effects are {\em primordial black holes}; see e.g.\ \cite{Carr} and references therein. It is remarkable that, very far from a black hole, the induced potential renormalizes boson masses by a negative constant. If many primordial black holes exist in the universe their cumulative mass renormalization effect could be quite large. 

\section{Robertson-Walker space-time}
\label{sec5}
The Robertson-Walker space-time describes the evolution of a homogenous isotropic universe. Its metric depends on a real parameter $K$ whose sign determines if the universe is open ($K>0$), flat ($K=0$), or closed ($K<0$).  All cases can be described by a line element of the form 
\begin{equation}
\label{ds2_RW} 
ds^2 = dt^2 - a(t)^2 \left( dr^2 + S(r)^2(d\theta^2 + \sin^2(\theta) d\varphi^2) \right)
\end{equation}
in coordinates $x=(t,r,\theta,\varphi)$, with $t$ the cosmological time, $a(t)$ the so-called {\em scale factor}, and 
\begin{equation}
S(r)\define \begin{cases} \sin(\sqrt{K}r)/\sqrt{K}& (K>0)\\ r& (K=0)\\ \sinh(\sqrt{|K|}r)/\sqrt{|K|} & (K<0)\end{cases} .
\end{equation} 
This space-time  can be naturally\footnote{We believe that there is no other embedding in five dimensional Minkowski space-time, but we are not aware of a proof of this in the literature.} embedded in $\mathbb{R}^{1,4}$ as follows (see \cite{Rosen}, B.5.), 
\begin{equation}
\label{embed_RW} 
\begin{split}
&Z^0 = \begin{cases} b(t)/\sqrt{K}& (K>0)\\ (1/2)\bigl( r^2/r_0+ r_0\bigr)a(t) + B(t) & (K=0)\\ a(t)C(r) & (K<0) \end{cases} \\
&Z^1 = \begin{cases} a(t) C(r) & (K>0)\\ (1/2)\bigl( r^2/r_0 - r_0\bigr)a(t) + B(t) & (K=0)\\ b(t)/\sqrt{|K|} & (K<0) \end{cases} \\
&Z^2 = a(t) S(r) \sin(\theta) \cos(\varphi)\\
&Z^3 = a(t) S(r) \sin(\theta) \sin(\varphi)\\ 
&Z^4 = a(t) S(r) \cos(\theta)
\end{split}
\end{equation}
with functions $b$ and $B$ defined by the following equations,  
\begin{equation}
\label{bdot}  
\dot{b}(t) = \sqrt{K+\dot a(t)^2} \quad (K\neq 0)
\end{equation} 
and
\begin{equation}
\dot B(t) = \frac1{2r_0 \dot a(t)} 
\end{equation} 
($\dot b(t)\equiv db(t)/dt$ etc.), and
\begin{equation}
C(r)\define \begin{cases} \cos(\sqrt{K} r)/\sqrt{K} & (K>0)\\ \cosh(\sqrt{|K|} r)/\sqrt{|K|} & (K<0)\end{cases} ;
\end{equation} 
$r_0>0$ is an arbitrary parameter. 

By straightforward computations we find the induced potential (see Appendix~\ref{appB3} for details) 
\begin{equation}
\label{RWpot}
V_{ind} = \frac{1}{4} \left( 6\frac{\ddot{a}}{a} + 3\frac{K+\dot{a}^2}{a^2} - \frac{\ddot{a}^2}{K+\dot{a}^2} \right) 
\end{equation}
for arbitrary $K$. Considering that the details of the embedding are different in the three cases $K>0$, $K=0$, and $K<0$, it is remarkable that the induced potential has such a simple form,  and, in particular, that it is continuous at $K=0$. 

We conclude this section with a preliminary discussion of the possible relevance of this induced potential in cosmology.  To simplify some formulas we only consider the case $K=0$. Inserting $H\define \dot a/a$ we find by simple computations 
\begin{equation}
V_{ind} = -\frac{H^2}{4}\Bigl( q^2+6q-3\Bigr) 
\end{equation} 
with $q\equiv \ddot a a/\dot a^2=-(\dot H+H^2)$ the usual {\em deceleration parameter}. Thus $V_{ind}\geq 0$ if the deceleration parameter is in the range $-6.46(4)\leq q\leq 0.464(1)$ and $<0$ otherwise, and $V_{ind}$ has typically the same order of magnitude as $H^2$. Since the value of the Hubble constant $H_0$ today corresponds to a  boson mass of approximately $10^{-33}\,$eV, this suggests to us that the induced potential is negligibly small in the universe at present and far back in time. To see if the induced potential could have an effect at early times we compute $V_{ind}$ for a universe with a scale factor growing as $a(t)=(t/t_*)^x$ for some exponent $x>0$ and some constant $t_*$, with $t$ the time after the big bang at $t=0$. We find 
\begin{equation}
V_{ind}=\frac{8x^2-4x-1}{4t^2}, 
\end{equation} 
i.e., as $t\to 0^+$, the induced potential diverges towards $-\infty$ for $0\leq x<(1+\sqrt{3})/4$ and towards $+\infty$ for $x>(1+\sqrt{3})/4$. This suggest that the induced potential could have had an important effect in the early universe.  In the next section we propose and study a self-contained model for the evolution of the early universe, taking into account the induced potential. As we will see, this model predicts that, as $t\to 0^+$, the scaling factor can vanish like $a(t)\to (t/t_*)^x$ with $x=(1+\sqrt{3})/4$ the critical value of the exponent, and such behavior is impossible without the induced potential. 

\section{Extrinsic curvature effects in the early universe} 
\label{sec6} 
Models of the early universe often assume a Robertson-Walker metric and matter represented by an isotropic Klein-Gordon field; see e.g.\ \cite{Linde,LythRiotto} for reviews or \cite{LythLiddle09} for a recent textbook on this topic. In this section we propose a generalization of such a model taking into account extrinsic curvature effects.  We also present results on the solution of this model and, to put them in perspective, compare them with results for the corresponding standard model where the external curvature effects are ignored. 

We obtain this model by adding our induced potential term to the action consisting of the usual Einstein-Hilbert term, cosmological term, and Klein-Gordon term with $\phi^4$-interaction \cite{LythLiddle09}:\footnote{In this section we write $\lambda$ and $m$ short for $\lambda_0$ and $m_0$, respectively. Moreover, the meanings of the symbols $x$, $y$, and $\phi_n$ are different from the ones in Section~\ref{sec2}.}
\begin{equation}
\label{S} 
S= \frac12\int d^4 x\, |g|^{1/2}\Bigl( \Mpl^2(R-2\Lambda) + g^{\mu\nu}(\partial_\mu\phi)(\partial_\nu\phi) - (m^2+ \chi V_{ind})\phi^2 -\frac{\lambda}2\phi^4\Bigr) 
\end{equation} 
with the Ricci scalar $R$ and the induced potential $V_{ind}$ in \Ref{Vind}. Our model parameters are the reduced Planck mass $\Mpl=1/\sqrt{8\pi G}$, the cosmological constant $\Lambda$, and the renormalized mass $m$ and coupling constant $\lambda$ of the Klein-Gordon field, respectively. To avoid writing similar formulas twice we use a parameter $\chi$ which is either $0$ (for the standard model case) or $1$ (for the extended model case with extrinsic curvature effects included). 

\subsection{Model details}
\label{RWmodels}
We are interested in a homogeneous Klein-Gordon field $\phi=\phi(t)$ (depending only on cosmological time $t$) on Robertson-Walker space-time with scaling factor $a(t)$. The standard method to  derive differential equations determining the time evolution of $a$ and $\phi$ in the case $\chi=0$ is as follows: restrict the Euler-Lagrange equations obtained from the action in \Ref{S} to the Robertson-Walker metric in \Ref{ds2_RW} and Klein-Gordon fields $\phi=\phi(t)$. Unfortunately we cannot use this method in the case $\chi=1$: this would require a formula describing how the induced potential changes with arbitrary variations of the metric, but this we do not have (since the induced potential depends on the embedding, and we do not know how to naturally change the embedding with the metric in general). 

We thus use an alternative method which, in the standard case $\chi=0$, leads to the same result as the method just described:  We insert the metric in \Ref{ds2_RW} and $\phi=\phi(t)$ into the action in \Ref{S}. This yields $S=const \int dt\, L$ (we can ignore the multiplicative constant) with  the Lagrangian 
\begin{equation}
\label{L} 
\begin{split} 
L= \Mpl^2(-3a\dot{a}^2-\Lambda a^3)  + \frac{a^3}2\Biggl(\dot\phi^2- \Bigl(m^2 + \frac{\chi}{4} \Bigl[ 6\frac{\ddot{a}}{a} + 3\frac{\dot{a}^2}{a^2} - \frac{\ddot{a}^2}{\dot{a}^2} \Bigr] \Bigr) \phi^2 -\frac{\lambda}{2}\phi^4 \Biggr)  
\end{split} 
\end{equation} 
where we used \Ref{RWpot}, setting $K=0$ for simplicity (we dropped total derivative terms; $\dot a\equiv da/dt$ etc). Note that, in the standard case $\chi=0$, the Lagrangian in \Ref{L} is of the following  usual type in dynamical systems: $L=L(a,\dot a,\phi,\dot\phi)$, corresponding to second order differential equations familiar from mechanics. However, in the case $\chi=1$, the Lagrangian also depends on $\ddot a$, and this leads to a 4-th order time evolution equation for $a$; see e.g.\ \cite{Simon} for some general background on such higher-derivative Lagrangian systems. We note that, even though the results described below are well-known in the case $\chi=0$, our method seems different from the ones used in the literature. 

By straightforward computations we obtain from \Ref{L} the following Euler-Lagrange equations, 
\begin{equation}
\label{ELphi}
\ddot \phi +3H\dot \phi +\Biggl(m^2 +\chi\Bigl[\dot H + 2H^2 -\frac{\dot H^2}{4 H^2}\Bigr]\Biggr)\phi+\lambda\phi^3=0
\end{equation}
and 
\begin{equation}
\label{ELa} 
\begin{split}
&2\dot H + 3H^2 -\Lambda + \frac1{\Mpl^{2}}\Bigl(\frac{\dot\phi^2}2-\frac{m^2}2\phi^2-\frac{\lambda}4\phi^4\Bigr) \\ &+\frac{\chi}{\Mpl^{2}}\Biggl( \phi^2\Bigl[ \frac{\ddot H}{2H}+\frac{\dot H^3}{4H^4}-\frac{H^2}2-\frac{\dot H\ddot H}{3H^3} + \frac{5\dot H}{12} -\frac{3\dot H^2}{8H^2}+\frac{\dddot H}{12 H^2}\Bigr]\\ &+\phi\dot\phi\Bigl[\frac{\dot H}{H}-\frac{2H}3 +\frac{\ddot H}{3H^2}-\frac{\dot H^2}{2H^3} \Bigr] +[\phi\ddot\phi+\dot\phi^2]\Bigl[\frac{\dot H}{6H^2}-\frac13\Bigr]\Biggr)=0
\end{split} 
\end{equation} 
with $H\equiv \dot a/a$. It is interesting to note that only $H$ and its derivatives appear in these equations, and thus the order of the Euler-Lagrange equation corresponding to $a$ is effectively reduced by one. It is important to note that the Lagrangian in \Ref{L} is invariant under time translations, and therefore the time evolution equations in \Ref{ELphi} and \Ref{ELa} allow for a conservation law given by \cite{Simon} 
\begin{equation}
\label{Idef} 
I \equiv \dot a\left( \frac{\partial L}{\partial \dot a} -\frac{d}{dt}\left(\frac{\partial L}{\partial \ddot a}\right)\right) + \ddot a\frac{\partial L}{\partial \ddot a} + \dot\phi \frac{\partial L}{\partial \dot \phi} -L
\end{equation} 
(using the Lagrange equations one easily checks that $dI/dt=0$). By straightforward computations we find that, up to a factor $a(t)^3$, this conservation law can be expressed in terms of $H$ and its derivatives as follows: 
\begin{equation}
\label{C}
\begin{split} 
\mathcal{I} \equiv -\frac{I}{3 a^3\Mpl^2}=& H^2 -\frac{\Lambda}{3}-\frac{1}{6\Mpl^2}\Bigl(\dot\phi^2 + m^2\phi^2 + \frac{\lambda}{2}\phi^4 \Bigr)\\ &+\frac{\chi}{\Mpl^{2}}\Bigl( \phi\dot\phi \Bigl[\frac16\frac{\dot H}{H}-\frac13H\Bigr]+\phi^2\Bigl[-\frac18\frac{\dot H^2}{H^2}+\frac1{12}\frac{\ddot H}{H} -\frac16H^2 +\frac14\dot H\Bigr] \Bigr)  .  
\end{split} 
\end{equation} 

In the standard case $\chi=0$, the two Friedmann equations are equivalent to \Ref{ELa} and $\mathcal{I}=0$: the dynamical system defined by the Lagrangian in \Ref{L} allows for many more solutions than Einstein's equations obtained by varying the action in \Ref{S}, but the additional solutions are eliminated by imposing that the constraint that the value of the conservation law in \Ref{Idef} is zero. {\em We assume that this is true also in the extended case $\chi=1$.} We emphasis that this is a hypothesis but, as we believe, a plausible one. To check it one should generalize the method described in the previous paragraph to the case $\chi=1$, but this we leave for future work. 

We thus propose the equations in \Ref{ELphi}, \Ref{ELa}, and $\mathcal{I}=0$ with $\mathcal{I}$ in \Ref{C} and for $\chi=1$, as extension of a standard cosmological model by extrinsic curvature effects. We are mainly interested in this model short after the big bang. We thus make the following ansatz for solutions,
\begin{equation} 
\label{ansatz} 
\begin{split}
\phi(t) &= t^{-y}\Bigl( \phi_0 + \phi_1 t + \phi_2 t^2 + \phi_3 t^3+\cdots\Bigr) \\
H(t) &= \frac{x}t + H_0 + H_1t + H_2t^2 + H_3t^3+\cdots . 
\end{split} 
\end{equation} 
We call solutions with $x=0$ {\em generic}, and the others {\em scaling solutions}. The reason for these names is as follows: since the equation of motions are invariant under time translations, one can replace in \Ref{ansatz} $t$ by $t-t_{0}$, $t_{0}$ arbitrary, and get another solution. Generically, the fields $a(t)$ and $\phi(t)$ at their derivatives at some time $t_0$ are finite, and the generic solution gives a series representing these fields in the vicinity of $t=t_0$ in terms of appropriate initial conditions at $t=t_0$. For scaling solutions, on the other hand, the time $t=0$ (say) is special: it is the time when the scaling factor vanishes like $(t/t_*)^x$ for some $t_*>0$, and, if $y>0$, the boson field diverges.

\subsection{Solutions}
\label{RWsolutions} 
We now discuss solutions of \Ref{ELphi}, \Ref{ELa}, and $\mathcal{I}=0$ with $\mathcal{I}$ in \Ref{C}, for the two cases $\chi=0$ and $\chi=1$ (some details on how we obtained these solutions are given in Appendix~\ref{appRW1}). 

For $\chi=0$, we found the following generic solution, 
\begin{equation}
\label{generic,  chi=0} 
\begin{split} 
\phi(t) &= \phi_0+\phi_1t-\frac{1}{2}(3H_0\phi_1+[m^2+\lambda \phi_0^2]\phi_0 )t^2+O(t^3)\\
H(t)&=H_0-\frac{\phi_1^2}{2\Mpl^2}t+\frac{\phi_1}{2\Mpl^2}(3H_0\phi_1+[m^2+\lambda \phi_0^2]\phi_0 )t^2+O(t^3) \\ 
H_0&=\sqrt{\Lambda/3 +\bigl( \phi_1^2/2 + m^2\phi_0^2/2 + \lambda\phi_0^4/4 \bigr)/(3\Mpl^2)}
\end{split} 
\end{equation} 
(the coefficients of the $O(t^3)$-terms are given in Appendix~\ref{appRW1}, \Ref{O3, chi=0}). This solution depends on two free parameters $\phi_0\equiv \phi(0)$ and $\phi_1\equiv \dot \phi(0)$, and this is the maximum possible number of free parameters: for $\chi=0$, \Ref{ELphi} and \Ref{ELa} are second- and first order differential equations, allowing for two- and one integration constants, respectively, and the constraint $\mathcal{I}=0$ reduces the number of free parameters by one. 

In the standard case $\chi=0$, we did not find any solutions as in \Ref{ansatz} with $x\neq 0$. However, in the extended case $\chi=1$, we found two such solutions: one with $x=3/4$, $y=1$ given by
\begin{equation}
\label{34}
\begin{split} 
\phi(t)=& \sqrt{\frac2{\lambda}}\Bigl( \frac1{4t} - \Bigl[\frac{2 m^2}{15}-\frac{126\lambda\Mpl^2}{65}  \Bigr] t + O(t^3) \Bigr) \\
H(t)=& \frac{3}{4t}  + \Bigl( \frac{m^2}{5}-\frac{594\lambda\Mpl^2}{65}\Bigr)t + O(t^3), 
\end{split} 
\end{equation} 
and another one with $x=(1+\sqrt{3})/4=0.683(0)$, $y=0$ given by\footnote{We write $0.683(0)$ for a numerical value $0.6830\pm 0.0001$.}
\begin{equation} 
\label{xc}
\begin{split}
\phi(t) =& \Mpl\Bigl(1.50(2) - \Bigl[0.228(0)\,{m}^{2} + 0.402(0)\,\lambda\Mpl^2 + 0.0883(5)\,{\Lambda}\Bigr]t^2 + O(t^4)  \Bigr) \\
H(t) = &\frac{0.683(0)}{t} - \Bigl( 0.0142(5)\,{m}^{2}+ 0.120(0)\lambda\Mpl^2 -  0.0690(4)\,\Lambda \Bigr)t + O(t^3) .
\end{split} 
\end{equation}
Note that the solution in \Ref{34} exists only if $\lambda>0$, whereas the solution in \Ref{xc} exists even if $\Lambda=\lambda=0$. Thus the latter solution is more robust and, as we believe, more interesting. We therefore give formulas for the coefficients of the $O(t^3)$-terms in \Ref{xc} in Appendix~\ref{appRW1}, \Ref{xcO3}. 

For $\chi=1$ we found the following generic solution,
\begin{equation}
\label{generic,  chi=1}
\begin{split} 
\phi(t)= & \phi_0+\phi_1t -\frac12\Bigl(3H_0\phi_1+\phi_0\Bigl[m^2+\lambda \phi_0^2 +H_1 + 2H_0^2 -\frac{H_1^2}{4H_0^2} \Bigr] \Bigr)t^2 + O(t^3) \\
H(t)= & H_0+H_1t+\Bigl( \Bigl[m^2 +\frac{\lambda}2\phi_0^2+H_0^2-\frac32 H_1\Bigr]H_0 +\frac{3H_1^2}{4H_0}\\ & +(2H_0^2-H_1)\frac{\phi_1}{\phi_0}+\bigl(\phi_1^2+2\Lambda\Mpl^2-6H_0^2\Mpl^2 \bigr)\frac{H_0}{\phi_0^2}\Bigr) t^2+O(t^3) 
\end{split} 
\end{equation} 
which depends on four free parameters: $\phi_0\neq 0$, $\phi_1$ as before and, in addition, $H_0\equiv H(0)$ and $H_1\equiv \dot H(0)$. Note that this is the maximum number of free parameters (since \Ref{ELa} now is third order, we have two more free parameters).

We note the {\em static} solutions, i.e.,  $\phi(t)=\phi_0$ and $H(t)=H_0$ independent of $t$, which are interesting special cases of the generic solutions above and exist if $m^2<0$, $\lambda>0$ ("Mexican hat potential") and $\Lambda>m^4/(4\lambda\Mpl^2)$. They are given by\footnote{Note that for every solution $\phi(t)$, $H(t)$ we give, $-\phi(t)$, $H(t)$ is also a solution.}
\begin{equation}
\label{chi=0, static} 
\phi_0 = \sqrt{-\frac{m^2}{\lambda}},\quad H_0=\sqrt{\frac{4\lambda\Lambda-m^4/\Mpl^2}{12\lambda}} \qquad\quad (\chi=0)
\end{equation}
and 
\begin{equation}
\label{chi=1, static}  
\phi_0 = \sqrt{\frac{2(-3m^2-2\Lambda)}{6\lambda +m^2/\Mpl^2}},\quad H_0=\sqrt{\frac{4\lambda\Lambda-m^4/\Mpl^2}{2(6\lambda +m^2/\Mpl^2)}} \qquad\quad (\chi=1)
\end{equation}
in the standard- and extended cases, respectively. Note that, for $\chi=1$, we have the additional conditions $\Lambda<-3m^2/3$ and $\lambda>-m^2/(6\Mpl^2)$ for these solutions to exist. Moreover, for $\Lambda>0$ and in both cases $\chi=0$ and $\chi=1$, there exists also the static solution with $\phi_0=0$ and $H_0=\sqrt{\Lambda/3}$. 

\subsection{Discussion} 
\label{RWdiscussion} 
As already mentioned, the generic solutions allows us to compute $a(t)$ and $\phi(t)$ in some interval around an arbitrary initial time $t=t_0$ from its initial conditions at $t=t_0$. In the standard case, they allow us to understand the importance of the so-called "slow roll" condition for inflation in this model \cite{LythLiddle09} in the following way: the equations of motion of the boson field $\phi(t)$ are identical with Newton's equations for a particle moving in one dimension under the influence  of a friction term and a potential $\mathcal{V}(\phi)=m^2\phi^2/2+\lambda\phi^4/4$: $\ddot\phi(t)=-3H(t)\dot\phi(t)-\mathcal{V}'(\phi(t))$, and \Ref{generic,  chi=0} implies $\dot H(t)=-\dot\phi(t)^2/(2\Mpl^2)$ and $\ddot H(t)=-6H(t)\dot H(t)+d{\mathcal{V}(\phi(t))}/dt$ (these equations can also be obtained directly from the Friedmann equations). This suggests that, starting from arbitrary initial conditions $\phi(0)$ and $\dot\phi(0)$ at some time $t=0$, $\phi(t)$ will increase with decreasing $\dot \phi(t)$ until $\dot\phi(t_1)=0$ at some time $t=t_1$, and, at this time, $\dot H(t_1)=\ddot H(t_1)=0$. In the vicinity of $t=t_1$,\footnote{We checked that the following result holds true for arbitrary differentiable boson potentials $\mathcal{V}(\phi)$.}
\begin{equation}
\label{series}
\begin{split} 
\phi(t)&=\phi_0 -\frac12\mathcal{V}'(\phi_0)(t-t_1)^2+O((t-t_1)^2)\\
\ln\Bigl( \frac{a(t)}{a(t_1)} \Bigr) &=H_0(t-t_1) - \frac{\mathcal{V}'(\phi_0)^2}{24\Mpl^2}(t-t_1)^4 + O((t-t_1)^5) 
\end{split} 
\end{equation} 
with $\phi_0\equiv \phi(t_1)$ and $H_0=\sqrt{\Lambda/3+\mathcal{V}(\phi_0)/(3\Mpl^2)}$, i.e., the time evolution of the scaling factor can be well approximated by the exponential law $a(t)\approx a(t_1)\exp\bigl( H(t_1)(t-t_1)\bigr)$, up to corrections that are negligible as long as  $\dot\phi(t)^2\ll 24\Mpl^2H(t_1)/|t-t_1|$.  We computed and plotted the approximation of the solution obtained by extending this series in \Ref{series} to higher orders and ignoring the $O((t-t_1)^n)$-terms, for different values of $n$, up to $n=8$. The result suggests that these series have a finite radius of convergence which, however, is often smaller than the full time interval where inflation occurs. 
 
 We now discuss the extended case $\chi=1$ and how it differs from the standard case $\chi=0$. Obviously, there are {\em two} essential differences: firstly, the generic solution in \Ref{generic,  chi=1} depends on {\em four} (rather than two for $\chi=0$) initial conditions, and secondly, there exist scaling solutions (which do not exist for $\chi=0$). In the next two paragraph we discuss these two differences and possible implications. 
  
 The generic solution in \Ref{generic,  chi=1} depending on four initial conditions is a consequence of the the Lagrangian depending also on $\ddot a(t)$ As illustrated in a simple toy model in Appendix~\ref{appToyModel}, even if such higher-order term in the Lagrangian is very small, it can lead to a much richer {\em qualitative} behavior of the system. Thus, at first sight, it seems that the model with $\chi=1$ is less predictive than the standard model with $\chi=0$.  However, our toy model also suggests that it is possible to restrict the initial conditions of the model with $\chi=1$ so as to obtain solutions that are similar to the ones for $\chi=0$. One natural way to impose such a restriction is suggested by the generic solution in  \Ref{generic,  chi=1}: obviously, this solution is not well-defined if $\phi_0=0$. Using translation invariance in time, we conclude that a generic solution such that $\phi(t_0)=0$ at some time $t=t_0$ can exist only if we impose some restrictions of the free parameters.  One can check that these restrictions on initial conditions at time $t=t_0$ are 
  \begin{equation}
  \label{H01}
  H_0^2=\frac{\Lambda}{3}+\frac{\phi_1^2}{6\Mpl^2},\quad H_1=-\frac{2\phi_1^2(\phi_1^2+2\Lambda\Mpl^2)}{3\Mpl^2(3\phi_1^2+4\Lambda\Mpl^2)}
  \end{equation} 
  with $H_0\equiv H(t_0)$ etc.\  (see Appendix~\ref{appRW1} for derivation). This leads to 
 \begin{equation}
 \label{phi0=0}
 \begin{split} 
\phi(t)=\phi_1 (t-t_0) -\frac{3}{2}\phi_1H_0(t-t_0)^2 + O((t-t_0)^3) , \quad 
 H(t) = H_0 + H_1 (t-t_0) \\ + \frac{4H_0^3 \phi_1^2(19\phi_1^4+36\phi_1^2\Mpl^2\Lambda+32(\Lambda\Mpl^2)^2)}{(3\phi_1^2+4\Lambda\Mpl^2)^2(5\phi_1^2+4\Lambda\Mpl^2))}(t-t_0)^2+O((t-t_0)^3) 
 \end{split} 
 \end{equation} 
 with $H_0$ and $H_1$ as in \Ref{H01}.  We thus propose the following hypothesis: {\em Only solutions that remain well-defined when $\phi(t)\to 0$ as $t\to t_0$, for some time $t_0$, can describe the evolution of the universe}.  This reduces the number of free parameters to two, which is equal to the number of free parameters in the standard case $\chi=0$: the time  $t=t_0$ where the boson field $\phi(t)$ vanishes, and the time derivative $\phi_1\equiv \dot\phi(t_0)$ of the boson field at that time. We believe that it is these restricted solutions of the extended case $\chi=1$ that behave, at larger times, similar to the ones on the standard case $\chi=0$. We note that there  might be also solutions where $\phi(t)>0$ for all times, but these we do not know how to restrict in a natural way.

We believe that the existence of scaling solutions in the extended case $\chi=1$, which do not have any analogue for $\chi=0$, could make the former model more predictive than the latter in the following way: for non-linear dynamical systems, scaling solutions often are attractors that capture the qualitative behavior of a large class of solutions (see e.g.\ \cite{Barenblatt}). As mentioned already above, the scaling solution in \Ref{xc} is more robust and thus, as we believe, more interesting than the one in \Ref{34}. We thus believe that it would be interesting to explore the validity of the following hypothesis: {\em Solutions that can describe the universe immediately after the big bang approach the scaling solution in \Ref{xc} as $t\to 0^+$.} Combining this hypothesis with the one above we get the following picture: at some time $t_0>0$, $\phi(t_0)=0$, and in some interval the time evolution is described by \Ref{H01}--\Ref{phi0=0}. The free parameters $\phi_1$ and $t_0$ of this solution are  then to be fixed such that, as $t\to 0^+$, this solution approaches the one in \Ref{xc}. For $t>t_0$, the boson field $\phi(t)$ grows with decreasing $\dot\phi(t)$ until $\dot\phi(t)=0$ at some time $t=t_1$, and close to this time the solution is, as we expect, similar to the one in \Ref{series}. If or not this picture holds true can be determined by a numerical solution, but this we leave to future work. We expect that the solution close to the time $t=t_1$ describes inflation, similarly as discussed after \Ref{series}. One important question is if such a solution can describe inflation (solving the horizon problem etc.\ \cite{LythLiddle09}) even in some time interval short after $t=0$ or not, i.e., if such a solution has one or two periods of inflation. At first sight it seems the answer to this question is negative:  $a(t)\to (t/t_*)^x$ as $t\to 0^+$ with $x=0.683(0)$, and previous work on power-law inflation \cite{AW,LM} suggests that inflation requires $x>1$. However,  in our scaling solution, $a(t)$ is well approximated by $(t/t_*)^x$ only in a small time interval, and for larger times the behavior is more complicated. We thus believe that this question is open.

\section{Conclusions}
\label{sec7} 
In this paper we proposed that the induced potential, which is well-established in constrained quantum mechanics, could also be relevant in brane-world scenarios where space-time is embedded in a higher dimensional ambient space. We showed that, by assuming that the propagating degrees of freedom are restricted to space-time by a strong confining potential, one finds that the Klein-Gordon equation on space-time is indeed modified by a induced potential term which does not depend of the details of the confining potential. While the formula for the induced potential we obtain is the same as in constrained quantum mechanics, the physical arguments to derive this potential are different. 

One striking feature of the induced potential is that it does not only depend on intrinsic- but also on extrinsic geometric properties of the embedded space-time. Thus, if the induced potential has observable consequences, it offers the intriguing possibility to test brane-world scenarios experimentally. As examples we computed the induced potential for Schwarzschild- and Robertson-Walker space-times. Our results suggest that, while the induced potential is usually negligibly small, it might be relevant in extreme situations like the early universe or in regions close to primordial black holes. We also proposed and studied a model for cosmological inflation with the effect of the induced potential included. At first sight this generalized model seems to be less predictive (since the generic solution of the equations of motion depend of four initial conditions, rather than two in the standard case). However, we found a natural condition reducing the number of initial conditions to two. Moreover, the extended model is different from the standard one also in that it allows for scaling solutions that do not depend on any free parameter. These scaling solution describe an expanding universe with a size approaching zero at some initial time. Our results  suggest that, in the extended model,  one only has to postulate a big bang, and the time evolutions of the universe is fixed, without much freedom to vary initial conditions. Anyway, it seems worthwhile to study this model further. 

For simplicity we restricted ourselves in this paper to the case where the ambient space is flat. It would be interesting to generalize our results in Section~\ref{sec2} to cases where the ambient space is curved. We expect that the induced potential thus obtained is the obvious generalization of one known in constrained quantum mechanics; see e.g.\ \cite{Mitchell}, Equation~(3.37). 

\appendix
\section{Derivation of the induced potential} 
\label{appA}
In this appendix we give details on how to derive the result stated in Section~\ref{sec2}. 

\subsection{Technicalities}
\label{appA1}
For compact manifolds $\mathcal{M}$ embedded in some ambient space, the existence of the coordinate system $(x,y)$ as described in the beginning of Section~\ref{sec2} is guaranteed by the so-called {\em Tubular Neigborhood Theorem}.  For this reason we believe that the assumption that such a coordinate system exists is not very restrictive. 

In the main text we argue that the details of the confining potential $V_{conf}$ are not important, and all that is needed is that this potential is such that the Schr\"odinger equation in \Ref{def chi} has a non-degenerate groundstate solution, obeying rather mild decay conditions, and with a finite energy gap to the first excited state. While this is plausible from a physics point of view, we emphasize that our derivation of the induced potential below can be easily promoted to  a mathematical proof only in the special case where the confining potential is (essentially) the box potential in \Ref{box} (the careful reader will find the spots where we implicitly make this assumption). It would be interesting to find a mathematical proof that  would apply to a much larger class of potentials. One technical challenge is to find arguments that avoid the use of local coordinates.

\subsection{Derivation}
\label{appA2} 
Taking the dot product of \Ref{alpha and A} with $\mathbf{t}_\nu$ and recalling
\Ref{t and n} we obtain 
\begin{equation}
\label{alpha}
\alpha_{k\mu\nu}\define \alpha_{k\mu}^{\phantom k\phantom\mu\lambda}g_{\lambda\nu} =  \mathbf{n}_k\cdot\partial_\mu\partial_\nu\mathbf{f} = \alpha_{k\nu\mu}
\end{equation}
which is useful in computations of the induced potential in examples. Similarly, by taking the dot product of \Ref{alpha and A} with $\mathbf{n}_j$, 
\begin{equation}
\label{A} 
A_{i\mu j}\define A_{i\mu}^{\phantom i\phantom\mu k}h_{kj} =
\mathbf{n}_j\cdot\partial_\mu\mathbf{n}_i = -A_{j\mu i}. 
\end{equation} 

We compute the action in \Ref{action2} by changing to the coordinates $\mathbf{\tilde{Z}}\equiv (x,y)$ in $\R^{1,n+p}$ defined in \Ref{tildeZ}, i.e., $\tilde{Z}^\mu=x^\mu$ for $\mu=0,\ldots,n$ and $\tilde{Z}^j=y^j$ for $j=n+1,\ldots,n+p$. We obtain  
\begin{equation}
\label{S1} 
S_0=\frac12\int d^{n+1}x\int d^py \, |G|^{1/2}\Bigl(G^{MN}\bigl(\tilde{\partial}_M\tilde\Phi\bigr)\bigl(\tilde{\partial}_N\tilde\Phi\bigr)-\bigl(m_{bare}^2+\tilde{V}_{conf}\bigr)\tilde{\Phi}^2 \Bigr) 
\end{equation} 
with $\tilde\Phi(x,y)=\Phi(\mathbf{\tilde{f}}(x,y))$, $\tilde{\partial}_M=\frac{\partial}{\partial\tilde{Z}^M}$, and $G_{MN}=\bigl(\tilde\partial_M\mathbf{\tilde{f}}\bigr)\cdot \bigl(\tilde\partial_N\mathbf{\tilde{f}}\bigr)$. Moreover, 
\begin{equation*}
(G_{MN}) \equiv \left( \begin{array}{cc} G_{\mu\nu} & G_{\mu j}\\ G_{i\nu} & G_{ij} \end{array} \right) = \left( \begin{array}{cc} \gamma_{\mu\nu}-G_{\mu k}h^{kl}G_{l\nu} & G_{\mu j}\\ G_{i\nu} & -h_{ij} \end{array} \right)
\end{equation*} 
with $G_{\mu j}= G_{j\mu} = y^k A_{k\mu j}$ and 
\begin{equation*}
\label{gamma}
\gamma_{\mu\nu} = g_{\mu\nu}-2y^k\alpha_{k\mu\nu} +y^ky^l
  \alpha_{k\mu\phantom\nu}^{\phantom
  k\phantom\mu\lambda}\alpha_{l\nu\lambda}^{\phantom l}. 
\end{equation*}
This allows us to compute the following, 
\begin{equation}
\label{det gamma} 
|\gamma|\define \bigl|\det(\gamma_{\mu\nu})\bigr| =
 |g|\Bigl(1-2y^k\alpha_{k\lambda}^{\phantom k\phantom\lambda\lambda} + y^k
 y^l\bigl(2\alpha_{k\lambda}^{\phantom
   k\phantom\lambda\lambda}\alpha_{l\sigma}^{\phantom k\phantom\sigma\sigma} -
 \alpha_{k\lambda}^{\phantom k\phantom\lambda\sigma}
 \alpha_{l\sigma}^{\phantom k\phantom\sigma\lambda} \bigr) + O(|y|^3) \Bigr).
\end{equation} 

Due to its special form it is possible to compute the determinant and the inverse of the matrix $(G_{MN})$ in a simple way as follows, $|G|=|\gamma||h|$
and 
\begin{equation*}
G^{MN} = \left( \begin{array}{cc} G^{\mu\nu} &
  h^{ik}G_{k\lambda}G^{\lambda\nu} \\
  G^{\mu\lambda}G_{\lambda k}h^{kj} & -h^{ij} + G_{k\lambda}G_{l\sigma}G^{\lambda\sigma}h^{ki}h^{lj} \end{array} \right)
\end{equation*}
with $\bigl(G^{\mu\nu}\bigr)$ the matrix inverse to $(\gamma_{\mu\nu})$. Thus
\begin{equation}
\label{GMN} 
\begin{split}
G^{\mu\nu} &= g^{\mu\nu} + O(|y|) \\ 
G^{\mu j} &= G^{j\mu}= y^k g^{\mu\lambda}A_{k\lambda}^{\phantom k\phantom\lambda j} + O(|y|^2) \\ 
G^{ij} &= -h^{ij} + y^ky^l A_{k\lambda}^{\phantom k\phantom\lambda i}
  A_{l\sigma}^{\phantom l\phantom\sigma j}g^{\lambda\sigma} + O(|y|^3). 
\end{split} 
\end{equation} 
It is straightforward to compute the higher order terms in $y^j$ but, as will be seen below, they are not needed. 

As mentioned in the introduction, it is important to rescale the Klein-Gordon fields so as to change the Jacobian $|G|^{1/2}$ in \Ref{S1} to $|g|^{1/2}|h|^{1/2}$. We thus introduce $\phi\define |\tilde\gamma|^{1/4}\tilde\Phi$ with $|\tilde\gamma|=|G|/(|g||h|)=|\gamma|/|g|$. Inserting this and $|\tilde\gamma|^{1/4}\tilde\partial_M\tilde\Phi=\tilde\partial_M\phi-(\tilde\partial_M|\tilde\gamma|)/(4|\tilde\gamma|)$
into \Ref{S1} and performing a partial integration yields 
\begin{equation}
\label{S2} 
S_0=\frac12\int d^{n+1}x\, |g|^{1/2}\int d^py\, |h|^{1/2} \Bigl( G^{MN}\bigl(\tilde{\partial}_M \phi\bigr)\bigl(\tilde{\partial}_N\phi\bigr)-\bigl(m_{bare}^2+\tilde{V}_{conf}+\tilde V_{ind} \bigr)\phi^2 \Bigr) 
\end{equation}
with
\begin{equation*}
\tilde V_{ind}= \frac{1}{4|\tilde\gamma|}\Biggl( G^{MN}\Bigl[\frac{3}{4|\tilde\gamma|}(\tilde\partial_M|\tilde\gamma|)(\tilde\partial_N|\tilde\gamma|) -(\tilde\partial_M\tilde\partial_N|\tilde\gamma|)\Bigr]-|g|^{-1/2}(\tilde\partial_M|\tilde\gamma|)\tilde\partial_N\bigl(|g|^{1/2}G^{MN}\bigr)\Biggr). 
\end{equation*}   
Using \Ref{det gamma} and \Ref{GMN} we find
\begin{equation}
\label{tilde Vind} 
\tilde V_{ind}(x,y)=V_{ind}(x)+O(|y|)
\end{equation}
with $V_{ind}(x)$ in \Ref{Vind}.

We now can insert the expansion $\phi(x,y)=\sum_\alpha \phi_\alpha(x)\tilde\chi_\alpha(y)$ with $\tilde\chi_\alpha(y)=|\epsilon|^{-p/2}\chi_\alpha(y/\epsilon)$ and $\chi_\alpha(y)$ the eigenfunctions defined by \Ref{def chi} and \Ref{def chi 1}. Note that the scaling is such that the $\tilde\chi_\alpha(y)$ are orthonormal. In the following we compute the different contributions to the integrand in
\Ref{S2}. In our computations we encounter integrals 
\begin{equation*}
\int d^p y\, |h|^{1/2} y^{k_1}\cdots y^{k_n} (\partial_i^a \tilde\chi_\alpha(y))(\partial_j^b \tilde\chi_\beta(y)) =
O(\epsilon^{n-a-b})
\end{equation*} 
for $a,b=0,1$. Thus, for fixed $a+b$, we only need to take into account the terms with $n\leq a+b$. This explains our truncations in \Ref{GMN}. 

Using \Ref{GMN} we obtain
\begin{equation*}
\label{L1}
\int d^p y\, |h|^{1/2}
G^{\mu\nu}(\partial_\mu\phi)(\partial_\nu\phi) = \sum_\alpha
g^{\mu\nu}(\partial_\mu\phi_\alpha)(\partial_\nu\phi_\alpha) + O(\epsilon)
\end{equation*}
and 
\begin{equation*}
\label{L2} 
2 \int d^p y\, |h|^{1/2} G^{\mu j }(\partial_\mu\phi)(\partial_j\phi) =  2\sum_{\alpha,\beta}  g^{\mu\lambda} (\partial_\mu\phi_\alpha) C_{\lambda \alpha\beta}\phi_\beta + O(\epsilon) 
\end{equation*}
with $C_{\lambda\alpha\beta}$ in \Ref{Cdef}. The following integral is computed using \Ref{def chi},  
\begin{equation*}
\label{L3}
\int d^p y\, |h|^{1/2}\Bigl(
-h^{ij}(\partial_i\phi)(\partial\phi_j)-(m_{bare}^2+\tilde V_{conf} )\phi^2\Bigr) = -\sum_\alpha (m_{bare}^2 + \mu_\alpha/\epsilon^2)\phi_\alpha^2
+O(\epsilon) 
\end{equation*}
with the $O(\epsilon)$-term coming from the $O(|y|)$-correction in \Ref{Vconf}. Using \Ref{GMN} and inserting $1=\sum_\gamma \int d^p y' |h|^{1/2}
\chi_\gamma(y)\chi_\gamma(y')$ yields 
\begin{equation*}
\label{L4}
\int d^p y\, |h|^{1/2}\bigl(G^{ij}+h^{ij}\bigr) (\partial_i\phi)(\partial_j\phi) = \sum_{\alpha,\beta,\gamma} g^{\mu\nu}C_{\mu\gamma\alpha}C_{\nu\gamma\beta}\phi_\alpha\phi_\beta + O(\epsilon) 
\end{equation*}
with $C_{\mu\alpha\beta}$ in \Ref{Cdef}. Finally, recalling \Ref{tilde Vind},  
\begin{equation*}
\label{L5} 
-\int d^p y\, |h|^{1/2} \tilde V_{ind}\phi^2 = -V_{ind}\sum_\alpha\phi_\alpha^2 + O(\epsilon) . 
\end{equation*} 
Collecting all integrals above we obtain the result in \Ref{action2}--\Ref{Vind}.

\section{Computation details}
\label{appB} 
\subsection{Schwarzschild black hole}
\label{appB1}
It is convenient to introduce normalized vectors $\mathbf{e}_\mu$ that allows to write the embedding in \Ref{embed_Schwarzschild} as 
\begin{equation*}
\mathbf{Z}=\mathbf{f}(x)\equiv 2r_s|1-r_s/r|^{1/2}\mathbf{e}_v+g(r)\mathbf{e}_2+r\mathbf{e}_r
\end{equation*}
for $x=(t,r,\theta,\varphi)$, i.e., 
\begin{equation*}
\begin{split} 
\mathbf{e}_v&=\begin{cases} \bigl( \sinh(t/[2r_s]),\cosh(t/[2r_s]) , 0,0,0,0\bigr) & (r >r_s) \\ 
\bigl( \cosh(t/[2r_s]),\sinh(t/[2r_s]) , 0,0,0,0\bigr) &(0<r<r_s)\end{cases} \\
\mathbf{e}_2&=(0,0,1,0,0,0)\\ 
\mathbf{e}_r&=(0,0,0,\sin(\theta)\cos(\varphi),\sin(\theta)\sin(\varphi),\cos(\theta)).
\end{split} 
\end{equation*} 
We find the following vectors normal to all tangent vectors and such that $h_{jk}=\delta_{jk}$ for $j,k=4,5$,
\begin{equation*}
\begin{split} 
\mathbf{n}_4&=-g'(r)|r/r_s-1|^{1/2}\mathbf{e}_v+(r_s/r)^{3/2}\mathbf{e}_2\\
\mathbf{n}_5&=(r_s/r)^{3/2}|1-r_s/r|^{1/2}\mathbf{e}_v+g'(r)(1-r_s/r)(r/r_s)^{1/2}\mathbf{e}_2-(r_s/r)^{1/2}\mathbf{e}_r
\end{split} 
\end{equation*} 
with $g'(r)$ in \Ref{gprime}. By straightforward computations we find the following non-zero components of $\alpha_{i\mu}^{\phantom i\phantom \mu\nu}$,
\begin{equation*}
\begin{split} 
\alpha_{40}^{\phantom 4\phantom 0 0} &=(r^2+rr_s+r_s^2)^{1/2}/(2rr_s),\quad 
\alpha_{41}^{\phantom 4\phantom 1 1} = 3r_s^2(r^2+rr_s+r_s^2)^{-1/2}/(2r^2)\\ 
\alpha_{50}^{\phantom 5\phantom 0 0}&=\alpha_{51}^{\phantom 5\phantom 1 1}= - (r_s/r)^{3/2}/(2r_s),\quad \alpha_{53}^{\phantom 5\phantom 3 3}=\alpha_{54}^{\phantom 5\phantom 4 4}= (r_s/r)^{3/2}/r_s 
\end{split} 
\end{equation*} 
(unless indicated otherwise, formulas  hold true for $0<r<\infty$) and the result in \Ref{Schwarzschild_result1}.

\subsection{Robertson-Walker space-time}
\label{appB3}
We first discuss the case $K>0$. Similarly as in Appendix~\ref{appB1} it is convenient to use normalized vectors $\mathbf{e}_{\mu}$, $\mu=0,1,r$, that allow to write the embedding in \Ref{embed_RW} as 
\begin{equation*} 
\mathbf{Z}= \mathbf{f}(x)\equiv \frac1{\sqrt{K}} b(t)\mathbf{e}_0 +a(t)\Bigl( C(r)\mathbf{e}_1+S(r)\mathbf{e}_r\Bigr)
\end{equation*} 
for $x=(t,r,\theta,\varphi)$. A normal vector to all tangent vectors and such that $h_{44}=1$ is 
\begin{equation*}
\mathbf{n}_4 = \frac{\dot{a}(t)}{\sqrt{K}}\mathbf{e}_0+\dot{b}(t)\Bigl( C(r) \mathbf{e}_1 +S(r)\mathbf{e}_r\Bigr).
\end{equation*} 
By straightforward computations we find the following non-zero components of $\alpha_{i\mu}^{\phantom i\phantom \mu\nu}$, 
\begin{equation*}
\alpha_{40}^{\phantom{40}0} = -\frac{\ddot a}{\dot b} , \quad \alpha_{41}^{\phantom{41}1} =\alpha_{42}^{\phantom{42}2} =\alpha_{43}^{\phantom{43}3} = - \frac{\dot{b}}{a}
\end{equation*}
and the result in \Ref{RWpot}. 

The computation for $K<0$ is similar, and the formulas above remain true with the following changes:  $\mathbf{e}_0$ and $\mathbf{e}_1$ are exchanged, and $\sqrt{K}$ is replaced by $\sqrt{|K|}$. 

For $K=0$ we find
\begin{equation*}
\mathbf{n}_4 = \frac12\Bigl[\Bigl(\frac{r^2}{r_0} + r_0\Bigr)\dot a(t) -\dot B(t) \Bigr]\mathbf{e}_0 +  \frac12\Bigl[\Bigl(\frac{r^2}{r_0} - r_0\Bigr)\dot a(t) -\dot B(t) \Bigr]\mathbf{e}_1 + r\dot a(t)\mathbf{e}_r
\end{equation*} 
and  $\alpha_{i\mu}^{\phantom i\phantom \mu\nu}$ as above with $\dot b=\dot a$. 

\subsection{Early universe models} 
\label{appRW1}
We give some details on how we obtained the results reported in Section~\ref{RWsolutions}. We used MAPLE to perform the computations described below. 

We denote the l.h.s.\ of \Ref{ELphi} and \Ref{ELa} as $F$ and $G$, respectively. Inserting the ansatz in \Ref{ansatz} we compute $F=t^{-s_F}\bigl(F_0 + F_1t + F_2t^2+\cdots \bigr)$ for some $s_F$ (determined by $x$ and $y$), and similarly for $G$. We then solve $F_n=G_n=0$ for $n=0,1,2,\cdots$. For low values of $n$ (often $n=0$) we obtain conditions allowing us to determine $x$, $y$ and the free parameters, and solving the equations for larger values of $n$ gives equations allowing to compute $\phi_n$ and $H_n$ in terms of the free parameters recursively. For the solutions thus obtained we compute the conservation law $I$ in \Ref{Idef} using \Ref{C}. This being constant is a useful check of our computations, and requiring this constant to be zero gives (usually\footnote{Two exceptions are discussed in the last paragraph of this section.}) an additional constraint on the free parameters of our solutions. 

In the standard case $\chi=0$ we find three different solutions of \Ref{ELphi} and \Ref{ELa}: (i) $x=4/3$, $y=1$, $\phi_0=\sqrt{2/\lambda}$; no free parameter, (ii) $x=2/3$, $y=0$; one free parameter $\phi_0$, (iii) $x=y=0$; three free parameters $\phi_0$, $\phi_1$, and $H_0$. Computing $I$ we find that the solutions (i) and (ii) are incompatible with $I=0$, and they therefore have to be discarded. For the solution (iii) the constraint $I=0$ determines $H_0$ in terms of $\phi_0$ and $\phi_1$, and we thus obtain the result given in \Ref{generic,  chi=0}. For completeness we also give the coefficients of the $O(t^3)$-terms of the solution (iii) in \Ref{generic,  chi=0}: 
\begin{equation}
\label{O3, chi=0} 
\begin{split} 
\phi_3= &\frac12(m^2+\lambda\phi_0^2)H_0\phi_0 + \frac{\phi_1^3}{2\Mpl^2} +\Bigl(\frac{\lambda}{8\Mpl^2}\phi_0^4 + \Bigl[\frac{m^2}{4\Mpl^2}-\frac{\lambda}{2}\Bigr]\phi_0^2 -\frac{m^2}{6}+\frac{\Lambda}{2} \Bigr) \phi_1 \\
H_3 =&  -\frac{31}{2\Mpl^2}(m^2+\lambda\phi_0^2)H_0\phi_0\phi_1 -\frac{3}{4\Mpl^4}\phi_1^4 
-\frac{1}{6\Mpl^2}\bigl(m^2 + \lambda\phi_0^2\bigr)^2 \phi_0^2 \\& - \frac1{\Mpl^2}\Bigl(\frac{\lambda}{4\Mpl^2}\phi_0^4 + \Bigl[\frac{m^2}{2\Mpl^2}-\frac{\lambda}{2}\Bigr]\phi_0^2 -\frac{m^2}{6}+ \Lambda \Bigr)\phi_1 \end{split}
\end{equation} 

In the extended case $\chi=1$ we obtain six different solutions: (i) $x=3/4$, $y=1$, $\phi_0=1/(2\sqrt{2\lambda})$; no free parameter, (ii) $x=4/3$, $y=1$, $\phi_0=1/(6\sqrt{\lambda})$; no free parameter, (iii) $x=(1+\sqrt{3})/4=0.683(0)$, $y=0$, $\phi_0=\sqrt{6(\sqrt{3}-1)/(21-11\sqrt{3})}\Mpl = 1.50(2)\Mpl$; no free parameter, (iv) $x=1$, $y=1/2$; one free parameter $\phi_0$, (v) $x=y=0$; five free parameters $\phi_0$, $\phi_1$, $H_0$, $H_1$, and $H_2$, (vi) $x=1$, $y=1/2$; one free parameter $\phi_0\neq 0$. Computing $I$ we find that the solutions (ii), (iv) and (vi) are incompatible with $I=0$, and they therefore have to be discarded. For the solutions (i) and (iii) we find that $\mathcal{I}$ in \Ref{C} is identically zero, and these solutions are therefore allowed. By straightforward computations we find the results in \Ref{34} and \Ref{xc} for the solutions (i) and (iii), respectively. As discussed in Section~\ref{RWdiscussion}, we believe that the solution (iii) in \Ref{xc} is of particular interest, and we therefore give the coefficients of the $O(t^3)$-terms in \Ref{xc}: 
\begin{equation} 
\label{xcO3} 
\begin{split}
\phi_3=& \Mpl\Bigl( 0.103(3)\, \lambda^{2}\Mpl^4+ 0.000278(3)
\,{\Lambda}^{2}+ 0.00989(0)\,{m}^{4}\\ &+ 0.00618(6)\,{m}^{2}
\Lambda+ 0.0286(9)\,\Lambda\,\lambda\Mpl^2 +
 0.0766(3)\,m^{2}\lambda\Mpl^2 \Bigr) \\
H_3=&  0.00913(8)\,{m}^{2}\Lambda+ 0.00298(1)\,{m}^2
\lambda\Mpl^2 + 0.0109(9)\,\lambda^{2}\Mpl^4
\\ &+ 0.00607(8)\,{\Lambda}^{2}+ 0.0118(0)\,\Lambda\,
\lambda\Mpl^2- 0.000568(1)\,{m}^{4}.
\end{split} 
\end{equation} 
For the solution (v) we compute $I$ in \Ref{Idef} using \Ref{C}, and we check that this is a constant depending on the free parameters of this solution. Demanding $I=0$ determines the free parameter $H_2$ in terms of the other ones, and we thus obtain the solution in \Ref{generic,  chi=1}. It is straightforward to compute the coefficients of the $O(t^3)$-terms in \Ref{generic,  chi=1}, but the result is lengthy, and we therefore do not give it here. 

The solution in \Ref{H01}--\Ref{phi0=0} can be obtained by the same methods as the other solutions described above, using the ansatz in \Ref{ansatz} with $x=y=\phi_0=0$. Alternatively, one can obtain this solution by a suitable limit from \Ref{generic,  chi=1} (the result for $H_0$ is easy to understand from \Ref{generic,  chi=1}, but the result for $H_1$ is somewhat more subtle; note that the naive guess $H_1=2H_0^2$ is incorrect).

It is interesting to note that, if one finds a solution of \Ref{ELphi} and \Ref{ELa} with $x\neq 0$ and $2y$ an integer, this  solution can have non-zero $I$ in \Ref{Idef} only if $3x$ is an integer. Indeed, \Ref{C} shows that $\mathcal{I}$ is a Laurent series in $t$, and, since $a(t)^3$ is equal to $t^{3x}$ times a Laurent series in $t$ (this follows from the definition of $H(t)$ and \Ref{ansatz}), $I$ is equal to $t^{3x}$ times a Laurent series in $t$. Thus, if $3x$ is not an integer, $I$ can be independent of $t$ only if it is identically zero, and this is the case for our solutions (i) and (iii) in the case $\chi=1$. For all other solutions of \Ref{ELphi} and \Ref{ELa} with $x\neq 0$, $3x$ is an integer, and $I$ is a constant depending on the free parameters. It is remarkable that, in all these latter cases, we find that $I=0$ is incompatible with other constraints.

\subsection{Higher-derivative Lagrangian systems: a toy model}
\label{appToyModel} 
We consider a generalization of the harmonic oscillator defined by the following Lagrangian 
\begin{equation}
\label{L0} 
L_0=\frac{1}{2}\Bigl( \dot q^2 -\omega^2q^2-\alpha\ddot q^2)
\end{equation} 
with $\omega>0$ and $\alpha > 0$ constants and $q=q(t)$ the dynamical variable. The Euler-Lagrange equation of this system is the following 4-th order differential equation, $\ddot q+\omega^2q+\alpha\ddddot q=0$, and it has the following general solution
\begin{equation}
\begin{split} 
q(t) = c_1\cos(\omega_+ t) + c_2\sin(\omega_+t) + c_3\cos(\omega_- t) + c_4\sin(\omega_-t) \\
\end{split}
\end{equation} 
with four free parameters $c_j$ and $\omega_\pm = \bigl([1\pm (1-4\alpha\omega^2)^{1/2}]/(2\alpha)\bigr)^{1/2}$, i.e., 
\begin{equation} 
\omega_- = \omega +\frac12\omega^3\alpha+O(\alpha^2) ,\quad \omega_+ = \frac{1}{\sqrt{\alpha}}-\frac12\sqrt{\alpha}\omega^2+O(\alpha^{3/2}).
\end{equation} 
We thus see that the limit $\alpha\to 0^+$ of this system is delicate: the general solution describes a "slow" motion with angular frequency $\omega_-$, which is a slightly perturbed motion of the harmonic oscillator (i.e., the solution of the model with $\alpha=0$, which only depends on two free parameters $c_{1,2}$), but on top of it there is a "fast" oscillatory motion with angular frequency $\omega_+$ which diverges as $\alpha\to 0^+$. Thus the limit $\alpha\to 0^+$ in the model is only well-defined for solutions where this "fast" motion is absent, i.e.,  if $c_3=c_4=0$. We can restrict to this subset of solutions which have a well-defined limit by the following constraints on initial conditions,
\begin{equation}
\ddot q + \omega_-^2 q=0,\quad \dddot q + \omega_-^2 \dot q=0\quad \mbox{ at } t=0. 
\end{equation}

\noindent {\bf Acknowledgments.} One of us (E.L.) would like to  thank Ariel  Goobar,  Fawad Hassan, Andrew Liddle, L\'arus Thorlacius and Mats Wallin for helpful remarks. We were inspired by a talk of Chryssomalis Chryssomalakos at the ``XXVII Workshop on Geometric Methods in Physics'' in Bialowieza (July 2008). This work was supported by the Swedish Science Research Council~(VR) and the G\"oran Gustafsson Foundation.


\begin{thebibliography}{99}

\bibitem{Maartens} R. Maartens: Brane-world gravity, Living Rev. Rel. 7, 7 (2004). URL (cited on August 28, 2010): http://www.livingreviews.org/lrr-2004-7

\bibitem{Branes} M. Pavsic and V. Tapia: Resource letter on geometrical results for embeddings and branes, arXiv:gr-qc/0010045v2

\bibitem{Schrodinger} E. Schr\"odinger: \"Uber das Verh\"altnis der Heisenberg-Born-Jordanschen Quantenmechanik zu der meinen, Ann. der Phys. 384, 734 (1926)  

\bibitem{JensenKoppe} H. Jensen and H. Koppe: Quantum mechanics with constraints, Ann. Phys. (N.Y.) 63, 586 (1971)

\bibitem{Marcus} R.A. Marcus: On the analytical mechanics of chemical reactions. Quantum mechanics of linear collisions, J. Chem. Phys. 45, 4493 (1966) 

\bibitem{C1} R. C. T. da Costa: Quantum mechanics of a constrained particle,  Phys. Rev. A 23, 1982 (1981)

\bibitem{C2} R. C. T. da Costa: Constraints in quantum mechanics, Phys. Rev. A 25, 2893 (1982)

\bibitem{MD} P. Maraner and C. Destri: Geometry-induced Yang-Mills fields in constrained quantum mechanics, Mod. Phys. Lett. A 8, 9861 (1993)

\bibitem{Mitchell} K.A. Mitchell: Gauge fields and extrapotentials in constrained quantum systems, Phys. Rev. A 63, 042112 (2001) 

\bibitem{FH} B. Froese and I. Herbst: Realizing holonomic constraints in classical and quantum mechanics, Comm. Math. Phys. 220, 489 (2001)

\bibitem{SJ} P.C. Schuster and R.L. Jaffe: Quantum mechanics on manifolds embedded in Euclidean space, Ann. Phys. (NY) 307, 132 (2003)

\bibitem{Gol}  A. Golovnev: Canonical quantization of motion on submanifolds, Rep. Math. Phys. 64, 59 (2009)

\bibitem{WT} J. Wachsmuth and S. Teufel:  Constrained quantum systems as an adiabatic problem, 	Phys. Rev. A 82, 022112 (2010) 

\bibitem{ES} P. Exner and P. Seba: Bound states in curved quantum waveguides, J. Math. Phys. 30, 2574 (1989)

\bibitem{GJ} J. Goldstone and R.L. Jaffe: Bound states in twisting tubes, Phys. Rev. B 45, 14100 (1992)

\bibitem{CLMTY} J.P. Carini, J.T. Londergan, K. Mullen, D.P. Murdock: Bound states in waveguides and bent quantum wires. I. Applications to waveguide systems, Phys. Rev. B 55, 9842 (1997)


\bibitem{Rosen} J. Rosen:  Embedding of various relativistic Riemannian spaces in pseudo-Euclidean spaces, Rev. Mod. Phys. 37, 204 (1965)

\bibitem{Borner} G. B\"orner: The early universe. Facts and fiction, Springer (2003) 

\bibitem{LythLiddle09} D.H.  Lyth, A.R. Liddle: The primordial density perturbation: cosmology, inflation and the origin of structure, Cambridge Univ. Press (2009) 

\bibitem{K1} E. Kasner: The impossibility of Einstein fields immersed in flat space of five dimensions, Am. J. Math. 43, 126 (1921)

\bibitem{Fronsdal} C. Fronsdal: Completion and embedding of the Schwarzschild solution,  Phys. Rev. 116, 778 (1959)


\bibitem{Sassi} G. Sassi: Analytical derivation of the embedding of the Schwarzschild solution in a flat space-time,  Nuovo Cim. B 102, 511 (1988)

\bibitem{Carr} B. J Carr: Primordial black holes as a probe of cosmology and high energy physics, astro-ph/0310838v1


\bibitem{Linde} A.D. Linde: The inflationary universe, Rep. Prog. Phys. 47, 925 (1984) 

\bibitem{LythRiotto} D.H. Lyth, A. Riotto: Particle physics models of inflation and the cosmological density perturbation, Phys. Rep. 314, 1 (1999) 

\bibitem{Simon} J. Z. Simon:  Higher-derivative Lagrangians, nonlocality, problems, and solutions,
Phys. Rev. D 41, 3720 (1990) 

\bibitem{Barenblatt} G.I. Barenblatt: Scaling, self-similarity, and intermediate asymptotics, Cambridge Universtity Press (1996)

\bibitem{AW} L.F. Abbott, M.B. Wise: Constraints on generalized inflationary cosmologies, Nucl. Phys. B 244,  541 (1984) 

\bibitem{LM} F. Lucchin F and S. Matarrese:  Power law inflation, Phys. Rev. D 32, 1316 (1985) 


\end{thebibliography}
\end{document}